\documentclass{article}
\usepackage{spconf,amsmath,amssymb,bm,graphicx}
\usepackage{algorithm,algpseudocode}

\usepackage{enumitem}
\setlist{nosep, leftmargin=14pt}

\usepackage{mwe} 
\usepackage{hyperref}

\usepackage{colortbl}
\usepackage{booktabs}
\usepackage{array}
\usepackage{multirow}
\definecolor{lightgray}{gray}{0.9}
\newcommand{\pmvalue}[2]{${#1} \scriptstyle\pm {#2}$}
\newcommand{\bpmvalue}[2]{$\mathbf{#1 \scriptstyle\pm #2}$}

\newcommand{\bfx}{\mathbf{x}}
\newcommand{\bfy}{\mathbf{y}}
\newcommand{\bfI}{\mathbf{I}}
\newcommand{\bftheta}{\bm{\theta}}

\newcommand{\calN}{\mathcal{N}}

\newcommand{\FIX}[2]{%
#1
}

\title{Fast Controllable Diffusion Models for Undersampled MRI Reconstruction}
\name{Wei Jiang$^{1}$ \qquad Zhuang Xiong$^{1}$ \qquad Feng Liu$^{1}$ \qquad Nan Ye$^{2}$ \qquad Hongfu Sun$^{1}$ \thanks{Code is available at \href{https://github.com/ppn-paper/ppn}{https://github.com/ppn-paper/ppn}}}

\address{$^{1}$ School of Electrical Engineering and Computer Science, University of Queensland, QLD, AUS \\
$^{2}$ School of Mathematics and Physics, University of Queensland, QLD, AUS}
\begin{document}
%
\maketitle

\begin{abstract}
Supervised deep learning methods have shown promise in undersampled Magnetic Resonance Imaging (MRI) reconstruction, but their requirement for paired data limits their generalizability to the diverse MRI acquisition parameters. Recently, unsupervised controllable generative diffusion models have been applied to undersampled MRI reconstruction, without paired data or model retraining for different MRI acquisitions. However, diffusion models are generally slow in sampling and state-of-the-art acceleration techniques can lead to sub-optimal results when directly applied to the controllable generation process. This study introduces a new algorithm called Predictor-Projector-Noisor (PPN), which enhances and accelerates controllable generation of diffusion models for undersampled MRI reconstruction. Our results demonstrate that PPN produces high-fidelity MR images that conform to undersampled k-space measurements with significantly shorter reconstruction time than other controllable sampling methods. In addition, the unsupervised PPN accelerated diffusion models are adaptable to different MRI acquisition parameters, making them more practical for clinical use than supervised learning techniques.

\end{abstract}

\begin{keywords}
Undersampled MRI Reconstruction, Deep Learning, Diffusion Models, Fast Controllable Generation, Inverse Problems.
\end{keywords}
\section{Introduction}
\label{sec:intro}
Deep learning has emerged as a valuable tool for undersampled Magnetic Resonance Imaging (MRI) reconstruction \cite{hammernik2018learning,han2019k,lee2018deep,gao2021accelerating,peng2022towards,korkmaz2023self}. However, most of these approaches are supervised, relying on paired data to learn a mapping between undersampled k-space measurements and the fully-sampled MR images. Moreover, supervised methods have poor generalizability when applied to different undersampling conditions, making them less practical for clinical use.

To address these limitations, Song et al. \cite{song2021solving} developed a fully unsupervised method that can reconstruct MRI and Computed Tomography images from undersampled signal acquisitions with similar performance but significantly enhanced generalizability to supervised methods. The idea is to train unsupervised generative diffusion models \cite{ho2020denoising,nichol2021improved,song2020score} using the MR images only, and then use physical model to guide the sampling process to generate MR images consistent with the k-space measurements.
However, diffusion models in general come at the cost of slow image generation speed.

\begin{figure}
    \centering
    \includegraphics[width=8.6cm]{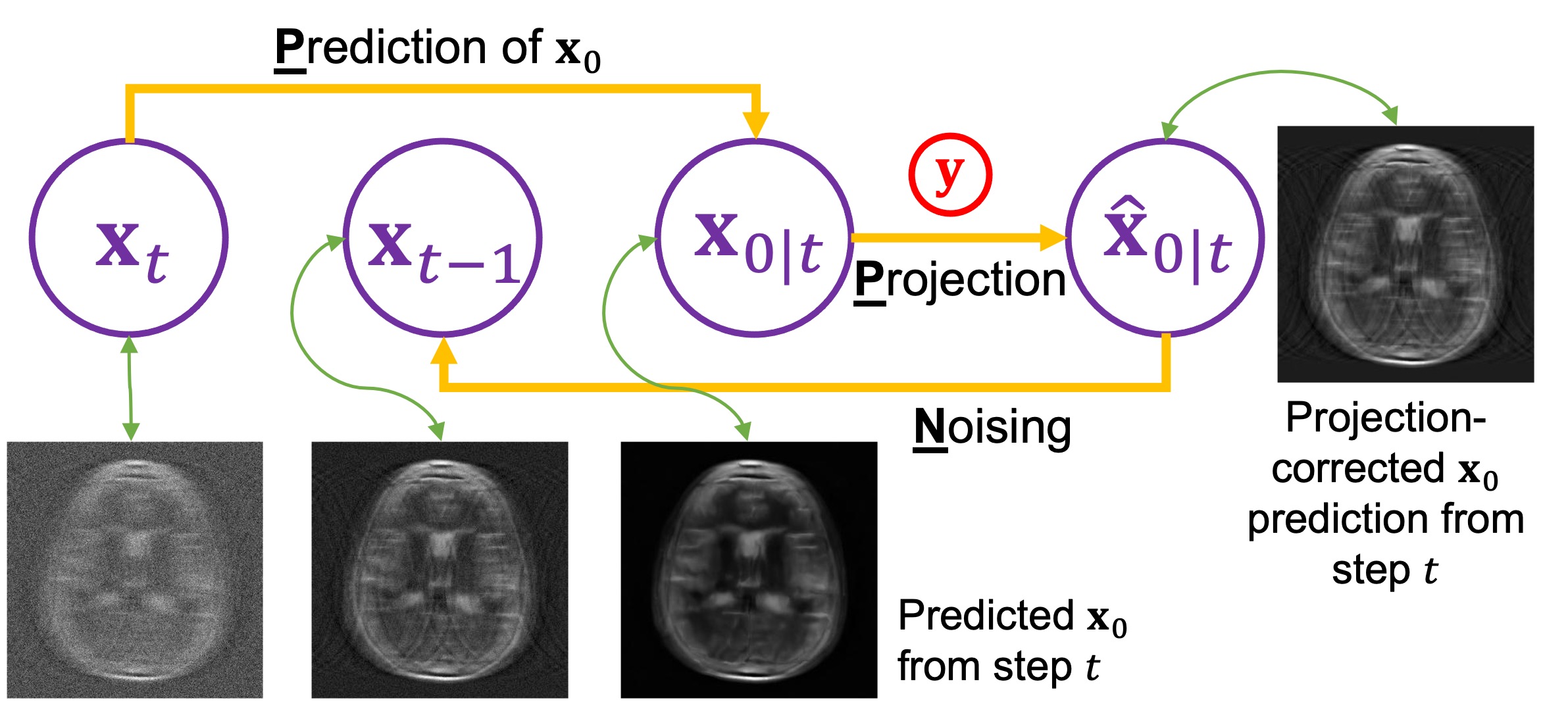}
    \caption{Illustration of a proposed PPN generation step.}
    \label{fig:onestep}
\end{figure}

Various methods have been proposed to accelerate the diffusion models, including Denoising Diffusion Implicit Models (DDIM) \cite{song2020denoising}. Although successful in speeding up unconditional image generation tasks, they lead to sub-optimal results when directly applied to accelerate the controllable generation process. This is because large reverse steps during the controllable sampling process can cause deviations from the manifolds \cite{chung2022dps}, particularly when projection algorithms are used to enforce k-space data consistency \cite{song2021solving,chung2022score}. It is well-known that sampling from out-of-distribution regions is challenging due to the lack of score information in the ambient space \cite{song2019generative}. Consequently, this deviation from the manifolds leads to an inefficient sampling trajectory, resulting in degraded reconstruction results.

We present a novel method called Predictor-Projector-Noisor (PPN) to accelerate the controllable generation process of diffusion models in undersampled MRI reconstruction. 
Our work follows the fully unsupervised framework of \cite{song2021solving}, but contributes a more efficient high-fidelity generative strategy. 
Specifically, during each generative step, the PPN method projects the DDIM-predicted noise-free image to match the k-space measurement and then noises it back to a higher noise level. This approach ensures the reverse trajectory stays within a chain of noisy manifolds converging to the data manifold, enabling fast generation without sacrificing image reconstruction accuracy.
Our experiments demonstrate that PPN outperforms other controllable generative processes in reconstructing high-fidelity MR images with significantly improved speed, and shows excellent generalizability across various under-sampling conditions without model retraining, due to their unsupervised learning nature.

\section{methodology}

\subsection{Preliminary: Diffusion Models}

Diffusion models (DMs) are a class of deep generative neural networks designed to sample from an unknown data distribution \(p_{\text{data}}\). They consist of two processes: a forward process that gradually perturbs inputs with noise to arrive at a tractable distribution (i.e. \({\cal N}({\bf 0},{\bf I})\)), and a reverse process that learns to undo the forward process by denoising. A well-known DM implementation is Denoising Diffusion Probabilistic Models (DDPM) \cite{ho2020denoising}, while Denoising Diffusion Implicit Models (DDIM) \cite{song2020denoising} accelerate the denoising process. 

\bigskip \noindent\textbf{DDPM}
The forward process of DDPM \cite{ho2020denoising} can be specified as a \(T\)-step Markov chain with a transition kernel \(q({\bf x}_{t}|{\bf x}_{t-1})={\cal N}\left({\bf x}_{t}|\sqrt{\alpha_{t}}{\bf x}_{t-1},(1-\alpha_{t}){\bf I}\right)\), where \({\bf x}_{0}\sim p_{\text{data}}\) and \(\alpha_t \in (0,1)\) for all \(t=1,\ldots,T\) are parameters chosen \cite{ho2020denoising} or trained \cite{nichol2021improved} to ensure that the distribution of \({\bf x}_T\) converges to \({\cal N}({\bf 0},{\bf I})\). Notably, diffused samples can be generated directly from a marginal distribution $q({\bf x}_{t}|{\bf x}_{0})={\cal N}\left({\bf x}_{t};\sqrt{\bar{\alpha}_{t}}{\bf x}_{0},(1-\bar{\alpha}_{t}){\bf I}\right)$, where \(\bar{\alpha}_{t}=\prod_{s=1}^{t}\alpha_{s}\). We can reparameterize it as:
\begin{equation}
	{\bf x}_{t}=\sqrt{\bar{\alpha}_{t}}{\bf x}_{0}+\sqrt{1-\bar{\alpha}_{t}}\boldsymbol{\epsilon}_t
 \label{eq:ddpm0}
\end{equation}
\noindent where \(\boldsymbol{\epsilon}_t\sim{\cal N}({\bf 0},{\bf I})\). The reverse process of DDPM utilizes a deep neural network, typically a U-Net, which is trained to predict the noise added at \({\bf x}_{0}\) given \({\bf x}_t\) and time \(t\). This predicted noise is denoted as \(\boldsymbol{\epsilon}_{\boldsymbol{\theta}}({\bf x}_{t},t)\). One denoising step in DDPM is:
\begin{equation}
	{\bf x}_{t-1}=\frac{1}{\sqrt{\alpha_{t}}}\left({\bf x}_{t}-\frac{1-\alpha_{t}}{\sqrt{1-\bar{\alpha}_{t}}}\boldsymbol{\epsilon}_{\boldsymbol{\theta}}({\bf x}_{t},t)\right)+\sqrt{1-\alpha_{t}}\boldsymbol{\epsilon}_{t}
 \label{eq:ddpm}
\end{equation}
\noindent\textbf{DDIM} DDIM \cite{song2020denoising} introduces a non-Markovian process that matching the marginal distribution \(q({\bf x}_{t}|{\bf x}_{0})\) to accelerate the denoising process, leading to a reformulation of \ref{eq:ddpm} as:
\begin{equation}
	{\bf x}_{t-1}=\sqrt{\bar{\alpha}_{t-1}}{\bf x}_{0|t}+\sqrt{1-\bar{\alpha}_{t-1}-\sigma_{t}^{2}}\boldsymbol{\epsilon}_{\boldsymbol{\theta}}({\bf x}_{t},t)+\sigma_{t}\boldsymbol{\epsilon}_{t}
 \label{eq:ddim}
\end{equation}
\noindent where \(\boldsymbol{\epsilon}_t\sim{\cal N}({\bf 0},{\bf I})\) and  
\begin{equation}
	{\bf x}_{0|t}=\frac{{\bf x}_{t}-\sqrt{1-\bar{\alpha}_{t}}\boldsymbol{\epsilon}_{\boldsymbol{\theta}}({\bf x}_{t},t)}{\sqrt{\bar{\alpha}_{t}}}
 \label{eq:ddim2}
\end{equation}
\noindent obtained by rewriting Eq. \ref{eq:ddpm0} with noise \(\boldsymbol{\epsilon}_{t}\) replaced by estimated \(\boldsymbol{\epsilon}_{\boldsymbol{\theta}}({\bf x}_{t},t)\). The parameter \(\sigma_{t}\) controls the noise magnitude in \(\boldsymbol{\epsilon}_{t}\), with DDIM being deterministic when \(\sigma_{t}=0\).

\subsection{MRI Inverse Problems}

In MRI inverse problems, we seek to reconstruct a signal \(\mathbf{x}\in \mathbb{R}^n\) from measurements \(\mathbf{y}\in \mathbb{R}^m = \mathbf{A}\mathbf{x} + \boldsymbol{\epsilon}\), where \(\boldsymbol{\epsilon}\in \mathbb{R}^m\sim \mathcal{N}(\mathbf{0},\sigma_{e}^{2}\mathbf{I})\) is noise, and \(\mathbf{A}\in \mathbb{R}^{m\times n} = \mathbf{M}\mathbf{F}\) represents the forward model with \(\mathbf{M}\) as the under-sampling mask and \(\mathbf{F}\) as the Fourier transform.

An inverse problem typically involves optimizing the cost function \( f(\tilde{\mathbf{x}}) = \frac{1}{2\sigma_{e}^{2}} \|\mathbf{y} - \mathbf{A}\tilde{\mathbf{x}}\|_{2}^{2} + \boldsymbol{s}(\tilde{\mathbf{x}}) \). This combines a fidelity term, ensuring consistency with measurements, with a prior as a penalty term to ensure \(\tilde{\mathbf{x}}\) looks natural. In this work, we use a pretrained diffusion model as the prior.

Tirer et al. \cite{tirer2018image} use the Plug-and-Play approach \cite{venkatakrishnan2013plug} to simplify inverse problem solving into the Iterative Denoising and Backward Projections (IDBP) algorithm. IDBP formulates the fidelity term as a projection of \(\tilde{\mathbf{x}}\) onto the affine subspace \(\{{\bf x} \in \mathbb{R}^{n} | {\bf A}{\bf x} = {\bf y}\}\), and its closed-form solution is \(\tilde{\mathbf{x}}' = {\bf A}^{\dagger}{\bf y} + ({\bf I} - {\bf A}^{\dagger}{\bf A})\tilde{\mathbf{x}}\) where \({\bf A}^{\dagger}\) is the \FIX{pseudoinverse}{write in the form of MRI inverse  problem} of \({\bf A}\). Using \(\mathbf{A}=\mathbf{M}\mathbf{F}\) and \(\mathbf{A}^{\dagger}=\mathbf{F}^{-1}\mathbf{M}\), we can rewrite the closed-form as:
\begin{equation}
	{\bf P}_{{\bf y}}(\tilde{\mathbf{x}})={\bf F}^{-1}({\bf M}{\bf y}+({\bf I}-{\bf M}){\bf F}\tilde{\mathbf{x}})
 \label{eq:ip}
\end{equation}
This projection-based approach is adopted in Song et al. \cite{song2021solving} (We will henceforth refer to it as 'MedScore') and DDNM \cite{wang2022zero}. The other category is gradient-based approach, including DPS \cite{chung2022dps}. In this paper, we compare our PPN with all three methods in section \ref{sec:ablation} and \ref{sec:results}.

\begin{algorithm}[H]
    \caption{Predictor-Projector-Noisor (PPN)}\label{alg:ppn}
    \begin{algorithmic}[1]
    \Require $S$, ${\bf y}$ \Comment{$S<T$\quad\ \ \ }
    \State {${\bf x}_{\text{zf}}\leftarrow{\bf F}^{-1}{\bf M}{\bf y}$}  \Comment{Zero-Filled}
    \State {${\bf x}_{S}\leftarrow\sqrt{\bar{\alpha}_{S}}{\bf x}_{\text{zf}}+\sqrt{1-\bar{\alpha}_{S}}\boldsymbol{\epsilon}_{S}$} \Comment{$\boldsymbol{\epsilon}_{S}\sim\calN(\bm{0},\bfI)$}
    \For{$t = S$ to $1$}                    
        \State {$\bfx_{0|t}\leftarrow\left(\bfx_t-\sqrt{1-\bar{\alpha}_t}\bm{\epsilon}_{\bftheta}(\bfx_t,t)\right)/\sqrt{\bar{\alpha}_t}$}  \Comment{\underline{P}rediction}

        \State {$\hat{\bfx}_{0|t}\leftarrow\mathbf{F}^{-1}(\mathbf{M}\bfy+(\bfI-\mathbf{M})\mathbf{F}\bfx_{0|t})$} 
        \Comment{\underline{P}rojection}

        \State {$\boldsymbol{\epsilon}_{t}\sim\calN(\bm{0},\bfI)$}
        \State {$\bfx_{t-1}\leftarrow\sqrt{\bar{\alpha}_{t-1}}\hat{\bfx}_{0|t}+\sqrt{1-\bar{\alpha}_{t-1}} \boldsymbol{\epsilon}_{t}$}
        \Comment{\underline{N}oisor\quad\ \ }
    
    \EndFor
    \State \Return {$\bfx_{0}$}
    \end{algorithmic}
    \end{algorithm}

\subsection{Predictor-Projector-Noisor (PPN) method}
\label{sec:ppn}
\begin{figure*}[!t]
    \centering
    \includegraphics[width=15cm]{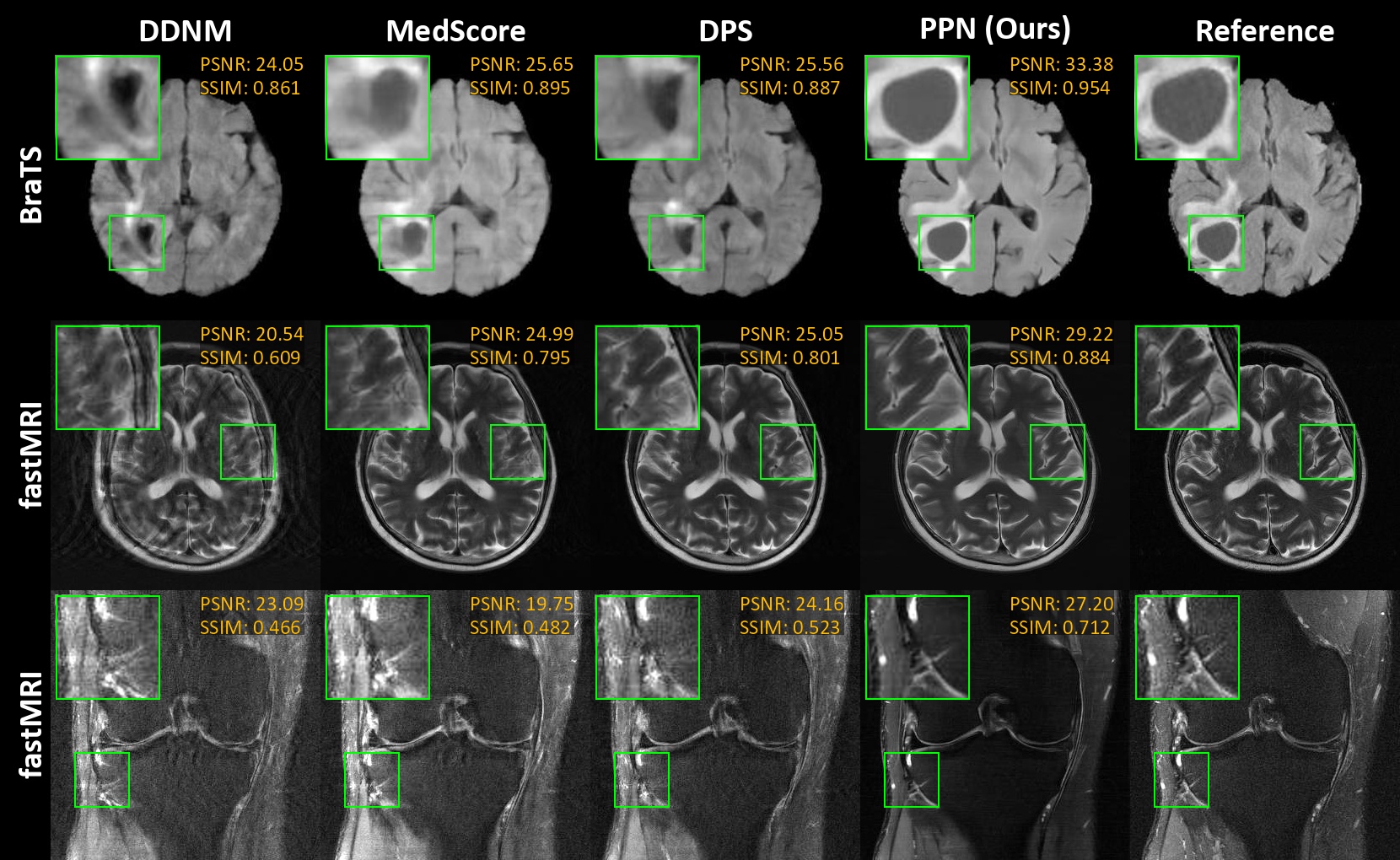} 
    \caption{MRI reconstructions for BraTS \cite{menze2014multimodal_brats1,bakas2017advancing_brats2} and fastMRI knee and brain \cite{zbontar2018fastMRI} at 8× acceleration, 50 NFEs.}
    \label{fig:image_compare}
\end{figure*}

Iterative generative methods for inverse problems depend heavily on accurate sampling information. However, the manifold hypothesis suggests that images exist on a lower-dimensional manifold embedded in a high-dimensional ambient space, posing a challenge. For example, gradient-based generation can fail due to the absence of scores in the ambient space \cite{song2019generative}. This issue is particularly prominent in projection-based methods such as DDNM \cite{wang2022zero} and MedScore \cite{song2021solving}, where early iterations often yield 'disharmonized' images outside the data manifold, a region that lacks denoising information.
Annealed Langevin Dynamics (ALD) \cite{song2019generative} and Diffusion Models (DMs) mitigate this by perturbing data with Gaussian noise, creating noisy manifolds. The Repaint method \cite{lugmayr2022repaint}, building on DMs, improves controllable generation using precise denoising information on noisy manifolds, reducing disharmony. However, its iterative approach increases neural function evaluations (NFEs), slowing the process.

Inspired by Repaint \cite{lugmayr2022repaint}, we propose a methood named Predictor-Projector-Noiser (PPN) for faster sampling in undersampled MRI reconstruction. Contrasting with Repaint's method of executing multiple iterations between two adjacent noise levels, our approach, based on deterministic DDIM \cite{song2020denoising}, increases the stride of each iteration while reducing the total number. As depicted in Figure \ref{fig:onestep} and outlined in Algorithm \ref{alg:ppn}, PPN loops once between \({\bf x}_t\) and \({\bf x}_{0|t}\) per iteration. A single step of conditional denoising is defined as follows:
\begin{equation}
    \mathbf{x}_{t-1}=\sqrt{\bar{\alpha}_{t-1}}\cdot{\bf P}_{{\bf y}}({\bf x}_{0|t})+\sqrt{1-\bar{\alpha}_{t-1}}\cdot\boldsymbol{\epsilon}_{t}
\label{eq:ppn}
\end{equation}
\noindent where \({\bf x}_{0|t}\) is obtained using Eq. \ref{eq:ddim2}, and \({\bf P}_{\bf y}(\cdot)\) is defined by Eq. \ref{eq:ip}. Starting with a noisy Zero-Filled sample \({\bf x}_{S}\), as supported by Fig. \ref{fig:ablation}, we can further accelerate PPN by iterating only through the final \(S\) steps, such as 50, out of the total 1,000 diffusion steps, significantly fewer in number.

\section{experiments}
\label{sec:experiments}
\begin{figure}
    \centering
    \includegraphics[width=7cm]{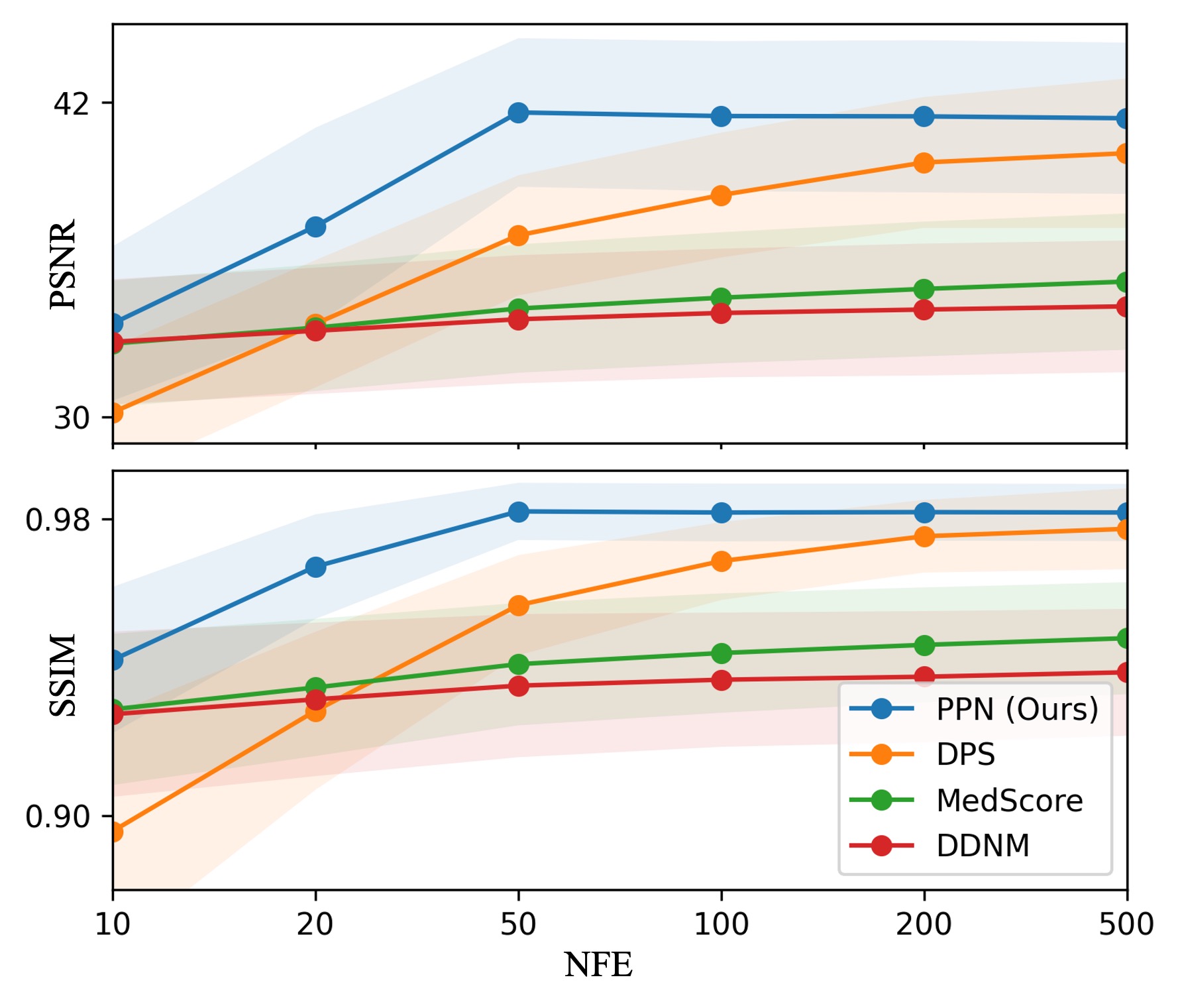}
    \caption{Performance vs. NFEs for 4× accel. reconstruction on BraTS \cite{menze2014multimodal_brats1,bakas2017advancing_brats2}. Shaded areas represent standard deviations.}
    \label{fig:ablation}
    \end{figure}

In this experiment section, our goal is to assess the PPN method's performance against three baselines: MedScore \cite{song2021solving}, DDNM \cite{wang2022zero}, and DPS \cite{chung2022dps}. We utilize the BraTS dataset \cite{menze2014multimodal_brats1,bakas2017advancing_brats2} for a standard retrospective study (Table \ref{tab:result}) and explore the effectiveness of NFEs on performance (Figure \ref{fig:ablation}). Additionally, we perform comparative evaluations with two other datasets, fastMRI Knee and Brain \cite{zbontar2018fastMRI} (Table \ref{tab:result2}).

\begin{table*}[t]
    \centering
    \caption{Results of undersampled MRI reconstructions on BraTS \cite{menze2014multimodal_brats1,bakas2017advancing_brats2} for 50 NFEs.}
    \vspace{5pt} 
    \begin{tabular}{c|cc|cc|cc}
        \toprule
        \multirow{2}{*}{Method}  & \multicolumn{2}{c|}{12$\times$  Acceleration } & \multicolumn{2}{c|}{8$\times$  Acceleration} & \multicolumn{2}{c}{4$\times$  Acceleration} \\
        \cmidrule(lr){2-3} \cmidrule(lr){4-5} \cmidrule(lr){6-7}
        &  PSNR$\uparrow$ & SSIM$\uparrow$ & PSNR$\uparrow$ & SSIM$\uparrow$ & PSNR$\uparrow$ & SSIM$\uparrow$ \\
        \midrule
        DDNM \cite{wang2022zero} &\pmvalue{27.42}{2.55} & \pmvalue{0.860}{0.037}&\pmvalue{30.38}{2.40} & \pmvalue{0.899}{0.027}&\pmvalue{33.74}{2.44} & \pmvalue{0.935}{0.019}       \\
        MedScore \cite{song2021solving}  & \pmvalue{27.91}{2.71} & \pmvalue{0.870}{0.036}&\pmvalue{31.47}{2.44} & \pmvalue{0.914}{0.023}&\pmvalue{34.14}{2.45} & \pmvalue{0.941}{0.017}        \\
        DPS \cite{chung2022dps}  & \bpmvalue{29.66}{3.06} & \pmvalue{0.892}{0.034}&\pmvalue{33.23}{2.41} & \pmvalue{0.928}{0.021}&\pmvalue{36.93}{2.29} & \pmvalue{0.957}{0.013}        \\
        \midrule
        \rowcolor{lightgray}
        PPN (Ours) &\pmvalue{29.07}{3.46} & \bpmvalue{0.902}{0.033}&\bpmvalue{37.51}{2.76} & \bpmvalue{0.964}{0.012}&\bpmvalue{41.62}{2.83} & \bpmvalue{0.982}{0.008}        \\
 
        \bottomrule
    \end{tabular} 
    \label{tab:result}
\end{table*}
\begin{table}[t]
    \centering
    \caption{SSIM scores for fastMRI knee and brain \cite{zbontar2018fastMRI} reconstructions at 8× acceleration with 50 NFEs.}
    \vspace{5pt} 
    \begin{tabular}{c|c|c}
        \toprule
        \multirow{1}{*}{Method}  & \multicolumn{1}{c|}{Knee} & \multicolumn{1}{c}{Brain} \\
        \midrule
        DDNM \cite{wang2022zero} &\pmvalue{0.675}{0.087} & \pmvalue{0.733}{0.060}  \\
        MedScore \cite{song2021solving}  & \pmvalue{0.719}{0.087} & \pmvalue{0.833}{0.049}    \\
        DPS \cite{chung2022dps}  & \pmvalue{0.709}{0.091} & \pmvalue{0.831}{0.071}  \\
        \midrule
        \rowcolor{lightgray}
        PPN (Ours) &\bpmvalue{0.827}{0.068} & \bpmvalue{0.918}{0.034}  \\
        \bottomrule
    \end{tabular} 
    \label{tab:result2}
\end{table}

\noindent\textbf{Dataset}
Our study utilized the Brain Tumor Segmentation (BraTS) 2021 dataset \cite{menze2014multimodal_brats1,bakas2017advancing_brats2}, converting 3D MRI volumes to 2D images (240x240 resolution), and used roughly 257,000 slices for training. We also used the fastMRI brain and knee datasets \cite{zbontar2018fastMRI}, selecting 'reconstruction\_rss' 2D images (320x320 resolution) and omitting the last 37.5\% of brain slices and the first 5 and last 3 knee slices, resulting in about 34,698 brain slices and 26,958 knee slices for training.

\noindent\textbf{Model Architecture}
We trained our models using an ablated diffusion model (ADM) \cite{dhariwal2021diffusion,nichol2021improved} with 1,000 steps, enabling learn\_sigma and adopting a cosine noise schedule as ADM suggests. For the BraTS datasets, we employed the U-Net architecture with 9.6M parameters, following the settings of MedScore \cite{song2021solving}. For the fastMRI knee and brain datasets, we used larger networks with the same 26.8M parameters.

\noindent\textbf{Training}
All models were trained on NVIDIA GPUs: Tesla V100 was used for BraTS with a learning rate of \( 2 \times 10^{-4} \) over 75 epochs, while A100 was used for the fastMRI brain dataset at \( 2 \times 10^{-4} \) for 317 epochs and for the fastMRI knee dataset at \( 5 \times 10^{-5} \) over 1,400 epochs.

\noindent\textbf{Baselines} 
For fair comparison, we re-implemented MedScore \cite{song2021solving}, DDNM's \cite{wang2022zero}, and DPS's \cite{chung2022dps} Algorithm 1 in ADM, testing at the same checkpoint and adjusting hyperparameters (\( \lambda=1.0 \) for MedScore, \( \zeta=10.0 \) for DPS). The Baselines' NFEs were configured using the DDIM setting, differing from PPN's approach as detailed in Section \ref{sec:ppn}.

Unless specified otherwise, we report  experiment results on 1,000 randomly selected test images using the peak signal-to-noise ratio (PSNR) and structural similarity index (SSIM) metrics, based on 50 NFEs and uniform 1D Cartesian masks.

\section{Ablation study}
\label{sec:ablation}
In this ablation study, we analyzed PPN's three main components. First, we replaced the noisor (Algorithm \ref{alg:ppn}, Line 7) with DDPM's Eq. 6 \cite{ho2020denoising}, resulting in DDNM's Algorithm 1 \cite{wang2022zero}. Second, we applied projection at \({\bf x}_t\) instead of \({\bf x}_0\), equivalent to MedScore's Algorithm 2 \cite{song2021solving}, by setting \(\lambda=1\) in Line 5. Third, we tested the effectiveness of NFEs, as illustrated in Figure \ref{fig:ablation}. Section \ref{sec:results} shows that PPN achieves optimal results using the final \(S=50\) steps across the majority of experiments on the three datasets in this study.

\section{Results} 
\label{sec:results}
Figure \ref{fig:image_compare} illustrates the reconstructions of various anatomies at 8x acceleration across all methods, showcasing PPN's fidelity to the ground truth. In Table \ref{tab:result}, PPN is compared with baseline methods for MRI undersampling at 4x, 8x, and 12x accelerations using 50 NFEs. Here, PPN outperforms the baseline methods at all accelerations except for the PSNR metric of the 12x acceleration, showcasing PPN's generalizability. This superiority is further supported by Table \ref{tab:result2}, where PPN achieves the highest SSIM for fastMRI knee and brain datasets at 50 NFEs and 8x acceleration. Additionally, Figure \ref{fig:ablation} shows PPN's performance against NFEs in PSNR and SSIM at 4x acceleration. Notably, PPN achieves better results with 50 NFEs compared to others at 500 NFEs.

\section{Conclusion}

We have introduced a novel approach called PPN for accelerating diffusion models in the context of undersampled MRI reconstruction. Our proposed method enables faster sampling while maintaining high fidelity, resulting in significant time reduction compared to other methods. Furthermore, our unsupervised model exhibits excellent generalizability and adaptability, enabling it to reconstruct various undersampling patterns without requiring retraining. Future work could focus on investigating PPN's performance in distribution shift tasks and its applicability to in-vivo scenarios.

\section*{Acknowledgment}

WJ acknowledges support from Dr F \& Mrs ME Zaccari Scholarship. HS acknowledges support from the Australian Research Council (DE210101297, DP230101628).

\section*{Compliance with Ethical Standards}

This research study was conducted retrospectively using human subject data made available in open access by Brain Tumor Segmentation (BraTS) \cite{menze2014multimodal_brats1,bakas2017advancing_brats2} and fastMRI \cite{zbontar2018fastMRI} datasets. Ethical approval was not required as confirmed by the license attached with the open access data.

\bibliographystyle{IEEEbib}
\bibliography{refs_shorter}

\end{document}